\documentclass[12pt,notoc]{JHEP3}
\usepackage{amsmath}
\usepackage{amssymb}
\newcommand{\sect}[1]{\setcounter{equation}{0}\section{#1}}
\def\rf#1{(\ref{eq:#1})}
\def\lab#1{\label{eq:#1}}

\newcommand{\beano}{\begin{eqnarray*}}
\newcommand{\enano}{\end{eqnarray*}}
\def\bea{\begin{eqnarray}}
\def\ena{\end{eqnarray}}

\relax

\font\fld=msbm10 at 12 pt

\newcommand{\fl}[1]{\mbox{\fld #1}}     

\def\bfm#1{\boldsymbol{#1}}
\def\ex#1{{\rm e}^{#1}}
\def\Buildrel#1\over#2\under#3{\mathrel{\mathop{\kern0pt #2}\limits^{#1}_{#3}}}
\def\bfm#1{\boldsymbol{#1}}
\def\alv{\bfm{\alpha}}

\def\lav{\bfm{\lambda}}
\newcommand\blank[1]{#1}
\renewcommand\blank[1]{}
\title{A T-duality interpretation of the relationship between massive and massless magnonic TBA systems}
\author{Patrick Dorey\\
Department of Mathematical Sciences\\ University of Durham\\ Durham DH1
3LE, UK.\\
e-mail: {\tt p.e.dorey@durham.ac.uk}}
\author{J.~Luis Miramontes\\
Departamento de F\'\i sica de Part\'\i culas, and\\ Instituto Gallego de
F\'\i sica de Altas Energ\'\i as (IGFAE),\\Universidad
de Santiago de Compostela\\ 15782 Santiago de Compostela, Spain.\\
e-mail: {\tt miramont@usc.es}}
\preprint{DCPT-06/27\\
hep-th/0609224
}
\abstract{
We propose an alternative understanding of the relationship between 
massive and massless magnonic TBA systems, using the T-duality
symmetries of the Homogeneous sine-Gordon models. 
This is shown to be in agreement with a previous treatment by
Dorey, Dunning and Tateo, based on the properties of Y-systems.
}
\begin{document}

\sect{Introduction.}
\label{Intro}

~\indent

The Thermodynamic Bethe ansatz (TBA)~\cite{ZAMTBA} is one of the most
effective ways to study the renormalization group (RG)
trajectories of two-dimensional integrable quantum field
theories. It enables the exact ground state energy on a
circle of circumference $R$ to be calculated from the solution of a
system of coupled non-linear integral equations, making it possible
to study the theory at all length scales, by varying the value of
$R$. The TBA equations can be deduced directly from the factorised
$S$-matrix of the theory, but this procedure becomes complicated when
the scattering is non-diagonal. This has often motivated an
alternative tactic, starting by the construction of a sensible 
set of TBA
equations, and investigating afterwards whether it corresponds to some
two-dimensional theory, typically defined by an action of the
form~\cite{ZAMPCFT}
\begin{equation}
S= S_{CFT} + \mu \int d^2x\> \Phi(x)\>.
\lab{Action}
\end{equation}
Here, $S_{CFT}$ denotes an action for a conformal field theory (CFT)
that governs the ultraviolet (UV) behaviour, $\mu$ is a dimensionful
coupling, and $\Phi$ is a perturbing operator.

Following this approach, a class of TBA systems whose structure is
encoded in a product of two simply-laced Dynkin diagrams was
constructed in~\cite{DKTBA} (see also~\cite{Martins:1991hi,RAVA}),
and they were
conjectured to describe a variety of integrable perturbed coset CFTs.
They include, as particular cases, many TBA systems previously
considered by other authors: in particular, those describing the 
well-studied perturbations
 of the unitary minimal models ${\cal M}_p$ for
$p=3,4,\ldots$ by their least relevant primary fields
$\Phi_{1,3}$~\cite{ZAMminimal,ZAMtric}.
In the construction of~\cite{DKTBA}, there is a particle and a TBA
equation for each node of the product diagram.
However, when the TBA system corresponds to a non-diagonal $S$-matrix,
some of these particles are fictitious and carry no energy and no
momentum.
They are called {\em magnons}, and it is common to refer to such TBA
systems as magnonic.
Magnonic TBA systems often admit massive and massless versions. In both
cases, they give rise to RG trajectories starting in the UV fixed
point specified by $S_{CFT}$. Then, either the trajectory flows to
some massive theory or it comes to another fixed point in the
infrared, depending on whether the system is massive or massless,
respectively.
In many cases the massive and massless versions of a given
magnonic TBA system correspond to the same action~\rf{Action} for
different signs of the coupling constant. This was originally noticed
in~\cite{ZAMtric}, where the TBA system for ${\cal M}_p +\mu
\Phi_{1,3}$ with $\mu>0$ was constructed. It is associated to
$A_{p-2}\times A_1$, and it turned out to be massless, in contrast to
the system that describes  the regime with $\mu<0$ which is
massive~\cite{ZAMminimal}. However, there are also cases where the
massive and massless TBA systems are {\em not}\/ related by continuation in
$\mu$. Examples are provided by the models $H_N^{(0)}$
(massive) and $H_N^{(\pi)}$ (massless) of~\cite{FATEEV}, which are
related by analytic continuation for $N$ odd, but not for $N$ even. In
the classification of~\cite{DKTBA}, they are associated to $D_N\times
A_1$.

This issue was clarified in~\cite{DDT}, where it was pointed out
the transformation could be understood directly in terms of the analytic
continuation of the corresponding TBA systems under
$\mu\rightarrow -\mu$. 
The massive and massless TBA systems corresponding to
the same product diagram are known to be related in a very simple way
that relies on the existence of a ${\fl Z}_2$ symmetry of the
associated Dynkin diagrams.
The authors of~\cite{DDT} pointed out that the continuation
$\mu\rightarrow -\mu$ will change a massive TBA system into a 
massless one
provided that the ${\fl Z}_2$ symmetry used in its construction
coincides with the ${\fl Z}_2$ symmetry that characterises the
periodicity properties of the associated Y-system. This systemetised
the previously-known zoology of examples in a simple rule, and also
provided a conceptual understanding, from the TBA point of view, of
why such a rule should exist.

The purpose of this letter is to propose an alternative understanding
of the relationship between the continuation $\mu\rightarrow -\mu$
and the transformation between massive and massless TBA systems that
does not rely on the properties of the Y-systems.
It will be deduced in the context of the Homogeneous sine-Gordon
(HSG) theories~\cite{HSG,QHSG,SMATHSG} by making use of Lagrangian
methods.
The TBA equations of the HSG theories~\cite{PATRICKTBA,SU3TBA} are
purely massive generalisations of the magnonic TBA systems
corresponding to products of the form $G\times A_{k-1}$. Moreover,
they admit a Lagrangian formulation in terms of perturbed gauged
Wess-Zumino-Witten (WZW) models where the required relationship
arises as a consequence of their target-space duality (T-duality)
symmetries~\cite{TDUAL}.

The letter is organised as follows. In section~\ref{Magnonic}, we
review the main features of the TBA systems constructed
in~\cite{DKTBA} and their associated Y-systems. For completeness,
section~\ref{TBA} explains the TBA argument of
\cite{DDT,DDTunpub}, relating the
continuation $\mu\rightarrow -\mu$ to the
transformation between massive and massless magnonic systems.
In section~\ref{HSGTBA}, we elucidate the relationship between
the TBA
systems of the HSG theories and the TBA systems
constructed in~\cite{DKTBA}.
The HSG TBA equations depend on a set of independent adjustable
parameters, and
their $\mu\rightarrow -\mu$ continuation is shown to be  equivalent
to a transformation among those parameters. Then, in
section~\ref{HSG}, we make use of the Lagrangian formulation of the
HSG theories to show that the same equivalence arises as a
consequence of T-duality. Moreover, this enables the resulting
transformation among the parameters to be written in terms of a
particular element of the Weyl group of $G$.
In the context of the HSG theories, the magnonic massive and massless
systems of~\cite{DKTBA} arise as the effective TBA systems describing
particular crossovers~\cite{PATRICKTBA}. Thus, this correspondence
points out a novel interpretation of the relationship between the
continuation $\mu\rightarrow-\mu$ and the transformation between
massive and massless TBA systems as a manifestation of T-duality,
which constitutes our main result.
Finally, section~\ref{Conclusions} contains our conclusions.

\sect{Magnonic TBA equations and Y-systems.}
\label{Magnonic}

~\indent
The structure of the magnonic TBA systems constructed in~\cite{DKTBA}
is encoded in the product $G\times H$ of two simply-laced Dynkin
diagrams $G$ and $H$. We will denote by $r_G$ and $r_H$ the ranks of
the corresponding algebras, and by $h_G$ and $h_H$ their Coxeter
numbers. For each node of the resulting product diagram, there is a
pseudoenergy $\varepsilon_a^i(\theta)$ and energy term
$\nu_a^i(\theta)$. Defining $L_a^i(\theta)=\ln\bigl(1+ {\rm
e}^{-\varepsilon_a^i(\theta)}\bigr)$, the system of TBA equations is
\begin{equation}
\nu_{a}^i(\theta)= \varepsilon_{a}^i (\theta)+
\sum_{b=1}^{r_H}\left(\phi_{ab}\ast L_{b}^i (\theta) -
\sum_{j=1}^{r_G}
\>G_{ij}\>
\psi_{ab}\ast L_{b}^j (\theta)
\right)\>,
\lab{TBAGen}
\end{equation}
for $i=1\ldots r_G$ and $a=1\ldots r_H$. In these equations, `$\ast$'
denotes the usual rapidity convolution $f\ast g(\theta) =
\int_{-\infty}^{+\infty} \frac{d\theta'}{2\pi}\> f(\theta-\theta')
g(\theta')$, and $G_{ij}$ is the incidence matrix of $G$. Similarly,
we will call $H_{ab}$ the incidence matrix of $H$.
The TBA kernels can be written as
\begin{equation}
\phi_{ab}=-i\frac{d}{d\theta}\ln S_{ab}^{\rm min}\>, \qquad
\psi_{ab}=-i\frac{d}{d\theta}\ln S_{ab}^{\rm F}\>
\end{equation}
in terms of the functions
\begin{equation}
S_{ab}^{\rm min}=\prod_{x\in A_{ab}} \{x\}\>, \qquad
S_{ab}^{\rm F}=\prod_{x\in A_{ab}}\bigl(x\bigr)\>,
\end{equation}
where $S_{ab}^{\rm min}$ are the minimal parts of the affine Toda
$S$-matrix elements corresponding to $H$. Here, $A_{ab}$ is a set of
integer numbers (possibly with repetitions),\footnote{For $H=A_k$,
this set is $A_{ab}=\bigl\{a+b+1-2l\mid l=1\ldots {\rm
min}(a,b)\bigr\}=
\bigl\{|a-b|+1\> \ldots \> a+b-1 , \>{\rm step}\;2\bigr\}$.} and the
basic blocks read
\begin{equation}
\{x\}= \bigl(x-1\bigr)\bigl(x+1\bigr)\>, \qquad
\bigl(x\bigr)\equiv\bigl(x\bigr)(\theta)= 
\frac{\sinh\frac{1}{2}\bigl(\theta+i\pi {x/ h_H}\bigr)}{ 
\sinh\frac{1}{2}\bigl(\theta-i\pi {x/ h_H}\bigr)}\>
\end{equation}
More details about the definitions of 
$S_{ab}^{\rm min}$ and $S_{ab}^{\rm F}$ can be found in~\cite{CORRDOR}.

The TBA expression for the ground state energy of the system 
on a circle of circumference $R$ is then
\begin{equation}
E_0(R)=E_{bulk}(M,R)- \pi\> c(r)/ 6R\>,
\end{equation}
where $c(r)$ is the so-called effective central charge, which
can be calculated in terms of the solutions to the TBA equations,
\begin{equation}
c(r)=\frac{3}{\pi^2}\> \sum_{i=1}^{r_G} \sum_{a=1}^{r_H} 
\int_{-\infty}^{+\infty} d\theta\> \nu_a^i(\theta)\> L_a^i(\theta)\>,
\lab{EffCC}
\end{equation}
and $E_{bulk}(M,R)$ is a bulk term. Here, $M$ is a mass scale, $r=MR$
a dimensionless overall scale, and the dependence on $R$ and on any
other mass scale in the theory enters via the energy terms
$\nu_a^i(\theta)$.

For each choice of $G\times H$, the authors of \cite{DKTBA}
defined $r_G$ different massive
magnonic TBA systems, one for each node $l=1\ldots r_G$ of the 
Dynkin diagram $G$, by choosing energy terms of the form
\begin{equation}
\nu_a^i(\theta)= \delta_{i,l}\> \mu_a\> r \cosh\theta\>,
\lab{EnTMassive}
\end{equation}
where $\mu_a$ are the components of the Perron-Frobenius eigenvector
of the Cartan matrix of $H$. The form of $\nu_a^i(\theta)$ reflects the
particle spectrum of the theory, which in this case consists of $r_H$
massive particles attached to the nodes $(l,a)$ for $a=1\ldots r_H$.
Their masses are given by $M_a^{(l)}= M\mu_a$, with
$M$ an overall mass scale. The particles that could be
associated to all the other nodes of $G\times H$ are magnons, and
they only contribute indirectly to $c(r)$ and $E_0(R)$, via
their effects on the non-magnonic pseudoenergies
$\varepsilon^l_a(\theta)$.

In contrast, as anticipated in the introduction, the massless magnonic 
systems of \cite{DKTBA} require the existence of a ${\fl Z}_2$
symmetry. Let us suppose that the Dynkin diagram $G$ has a ${\fl
Z}_2$ symmetry $\omega:G\rightarrow G$ that relates two nodes $l$ and
$l'=\omega(l)\not=l$. Then, we can associate a massless magnonic TBA
system to the node~$l$ by choosing the energy terms as follows:
\begin{equation}
\nu_a^i(\theta)= \delta_{i,l}\> \frac{1}{2}\> \mu_a r\> {\rm e}^{-\theta} + \delta_{i,\omega(l)}\> \frac{1}{2}\> \mu_a r\> {\rm e}^{+\theta}\>. 
\lab{EnTMassless}
\end{equation}
In this case, the particle spectrum of the theory consists of $2r_H$
massless particles: $r_H$ left-movers and $r_H$ right-movers
associated with the nodes $(l,a)$ and $\left(\omega(l),a\right)$,
respectively. This exhibits that $M$ is a crossover scale in this
case. Again, the particles associated to all the other nodes in
$G\times H$ are magnons.
All the massless TBA systems that have been discovered to
date are related to massive systems by means of a transformation of
the energy terms similar to the one that takes~\rf{EnTMassive}
into~\rf{EnTMassless}~\cite{DKTBA,DDT}, a process which can sometimes
be quite elaborate \cite{Dorey:1996ms}.

An important feature of the TBA equations is that they provide
($r$-dependent) solutions to a set of functional algebraic equations
called the Y-system~\cite{Ysystems}. The Y-system corresponding
to~\rf{TBAGen} is
\begin{equation}
Y_a^i\bigl(\theta+\frac{i\pi}{h_H}\bigr)\> 
Y_a^i\bigl(\theta-\frac{i\pi}{ h_H}\bigr)
=\prod_{b=1}^{r_H} \bigl(1+Y_b^i(\theta)\bigr)^{H_{ab}}
\prod_{j=1}^{r_G} \left(1+\frac{1}{Y_a^j(\theta)}\right)^{-G_{ij}},
\lab{YSystem}
\end{equation}
with $Y_a^i(\theta)={\rm e}^{\>\varepsilon_a^{i}(\theta)}$ an entire
function of $\theta$. Notice
that the Y-system is completely independent of the form of the energy
terms and, in particular, of the value of the dimensionless scale
$r$. The role of the energy terms is to fix the asymptotic behaviour
of $Y_a^i(\theta)$. Indeed, since $Y_a^i={\rm
e}^{\>\varepsilon_a^{i}}$, it is controlled by the asymptotic
behaviour of $\epsilon_a^i(\theta)$ that, in turn, is dominated by
the energy term $\nu_a^i(\theta)$.
In particular, the asymptotic behaviour of the solutions to the 
massive TBA system specified by~\rf{EnTMassive} is
\begin{equation}
Y_a^i(\theta)
\; \buildrel \theta\;\rightarrow \pm\infty \over{\hbox to
50pt{\rightarrowfill}}
\begin{cases}
\exp\Bigl(\frac{1}{2}\> \mu_a r \>\ex{\pm\theta}\Bigr)\approx
\exp\Bigl(\nu_a^l(\theta)\Bigr)& \text{for}\quad i=l, \\
\>y_a^i& \text{for}\quad i\not=l,
\end{cases}
\end{equation}
where $y_a^i$ are the solutions to the constant 
($\theta$-independent) Y-system
\begin{equation}
\left(y_a^i\right)^2
=\prod_{b=1}^{r_H} \bigl(1+y_b^i\bigr)^{H_{ab}}
\prod_{j=1}^{r_G} \left(1+\frac{1}{ y_a^j}\>\right)^{-G_{ij}}\>.
\lab{ConstantY}
\end{equation}
In contrast, the asymptotic behaviour of the solutions to the 
massless TBA system whose energy terms are~\rf{EnTMassless} is given by
\begin{eqnarray}
Y_a^i(\theta)
&&\hspace{-0.2cm}\buildrel \theta\;\rightarrow -\infty \over{\hbox to
50pt{\rightarrowfill}}
\begin{cases}
\exp\Bigl(\frac{1}{2}\> \mu_a r 
\>\ex{-\theta}\Bigr)& \text{for}\quad i=l, \nonumber\\
\>y_a^i& \text{for}\quad i\not=l,
\end{cases}\\[5pt]
&&\hspace{-0.2cm}\buildrel \theta\;\rightarrow +\infty \over{\hbox to
50pt{\rightarrowfill}}
\begin{cases}
\exp\Bigl(\frac{1}{2}\> \mu_a r \>\ex{+\theta}\Bigr)& \text{for}\quad i=
\omega(l), \\
\>y_a^i& \text{for}\quad i\not=\omega(l).
\end{cases}
\lab{MasslessAsymp}
\end{eqnarray}

It is important to stress that fixing the asymptotic behaviour of the
Y-functions is not quite enough to ensure that the solutions to the
Y-system~\rf{YSystem} correspond to solutions to the ground-state TBA
equations~\rf{TBAGen}. The reason is that a given Y-system admits
more solutions than those related to the original TBA equations. In
fact, the same Y-system describes different excited states of the
model, and the difference between the various excited state solutions
is in their analytical structure~\cite{NoZeroes,Excited,PATROB,PATROB2}.
The simplest case concerns the ground state itself, which provides the
solution to the original TBA system. It is recovered by restricting
ourselves to Y-functions which are free of zeroes 
in the strip $-\pi/h_H<{\rm Im\/}\bigl(\theta\bigr)
<\pi/h_H$. With this restriction, the
Y-system~\rf{YSystem} with appropriate asymptotic behaviour is
completely equivalent to the system of TBA equations~\rf{TBAGen}.

One of the main properties of Y-systems, first noticed
in~\cite{Ysystems}, is that they generate periodic functions. 
In our case the Y-functions satisfy~\cite{DKTBA,GliozziTateo}
\begin{equation}
Y_a^i(\theta+ i\pi\> P) =Y_{\>\overline{a}}^{\>\overline{\imath}}(\theta) 
\qquad \text{with}\qquad 
P= \frac{h_G+h_H}{h_H}\>,
\lab{PeriodY}
\end{equation}
where the nodes $\overline{\imath}$ and $\overline{a}$ are {\em
conjugate} to the nodes $i$ and $a$ on the Dynkin diagrams $H$ and
$G$, respectively, and conjugation acts on Dynkin diagrams in the
same way as charge conjugation acts on the particles in an affine
Toda field theory (see fig.~\ref{DynkinDiag}). 
The period $P$ can then be
related to the conformal dimension of the perturbing operator $\Phi$;
see~\cite{Ysystems,DKTBA} for more details.

Eq.~\rf{PeriodY} was originally verified by direct successive
substitutions in~\rf{YSystem} for particular (low rank) choices of
$G\times H$.
Subsequently, the periodicity for the cases of the form $G\times A_1$
was proved in~\cite{Aperiodicity} ($G=A_n$),~\cite{Dperiodicity}
($G=D_n$), and~\cite{Fomin}. Proofs for $A_m\times A_n$ with
$m,n\not=1$ have been recently provided in~\cite{Volkov}.

\newcommand{\sts}{\footnotesize}
\setlength{\unitlength}{1mm}
\thicklines
\newsavebox{\An}
\sbox{\An}{\begin{picture}(52,5)(0,-3.5)
\put(0,0){\mathversion{bold}$A_n$}
\multiput(10,0)(10,0){5}{\circle*{1.75}}
\multiput(10,0)(10,0){2}{\line(1,0){10}}
\multiput(31,0)(1,0){9}{\circle*{.2}}
\put(40,0){\line(1,0){10}}
\put(10,-4){\makebox(0,0)[b]{{\sts 1}}}
\put(20,-4){\makebox(0,0)[b]{{\sts 2}}}
\put(30,-4){\makebox(0,0)[b]{{\sts 3}}}
\put(40,-4){\makebox(0,0)[b]{{\sts {\em n}--1}}}
\put(50,-4){\makebox(0,0)[b]{{\sts {\em n}}}}
\end{picture}}
\newsavebox{\Dn}
\sbox{\Dn}{\begin{picture}(70,15)(0,-5)
\put(0,0){\mathversion{bold}$D_n$}
\multiput(10,0)(10,0){5}{\circle*{1.75}}
\multiput(10,0)(10,0){2}{\line(1,0){10}}
\multiput(31,0)(1,0){9}{\circle*{.2}}
\put(40,0){\line(0,1){10}}
\put(40,0){\line(1,0){10}}
\put(10,-4){\makebox(0,0)[b]{{\sts 1}}}
\put(20,-4){\makebox(0,0)[b]{{\sts 2}}}
\put(30,-4){\makebox(0,0)[b]{{\sts 3}}}
\put(40,-4){\makebox(0,0)[b]{{\sts {\em n}--2}}}
\put(40,10){\circle*{1.5}}
\put(43,9){\makebox(0,0)[b]{{\sts {\em n}}}}
\put(50,-4){\makebox(0,0)[b]{{\sts {\em n}--1}}}
\end{picture}}
\newsavebox{\En}
\sbox{\En}{\begin{picture}(70,15)(0,-5)
\put(0,0){\mathversion{bold}$E_n$}
\multiput(10,0)(10,0){5}{\circle*{1.75}}
\multiput(10,0)(10,0){1}{\line(1,0){10}}
\multiput(21,0)(1,0){9}{\circle*{.2}}
\multiput(30,0)(10,0){2}{\line(1,0){10}}
\put(30,0){\line(0,1){10}}
\put(30,10){\circle*{1.5}}
\put(10,-4){\makebox(0,0)[b]{{\sts 1}}}
\put(20,-4){\makebox(0,0)[b]{{\sts 2}}}
\put(30,-4){\makebox(0,0)[b]{{\sts{\em n}--3}}}
\put(40,-4){\makebox(0,0)[b]{{\sts{\em n}--2}}}
\put(50,-4){\makebox(0,0)[b]{{\sts{\em n}--1}}}
\put(33,9){\makebox(0,0)[b]{{\sts{\em n}}}}
\end{picture}}

\FIGURE[ht]
{
\begin{picture}(130,60)(0,30)
\put(0,80){\usebox{\An}}
\put(60,82.5){$\overline{a}=n+1-a$, $\quad a=1\ldots n$}
\put(0,57){\usebox{\Dn}}
\put(60,68){$n$ even:}\put(76,68){$\quad\overline{a}=a$, $\quad a=1\ldots n\quad$}
\put(60,61){$n$ odd:}\put(76,61){$\quad\overline{a}=a$, $\quad a=1\ldots n-2\quad$}
\put(76,54){$\quad\overline{n-1}=n$, $\quad\overline{n}=n-1$}
\put(0,33){\usebox{\En}}
\put(60,43){$n=6$:}\put(76,43){$\quad\overline{1}=5$, $\;\overline{2}=4$, $\;\overline{3}=3$, $\;\overline{6}=6$}
\put(60,36){$n=7,8$:}\put(76,36){$\quad\overline{a}=a$, $\quad a=1\ldots n\quad$}
\end{picture}
\caption{\small
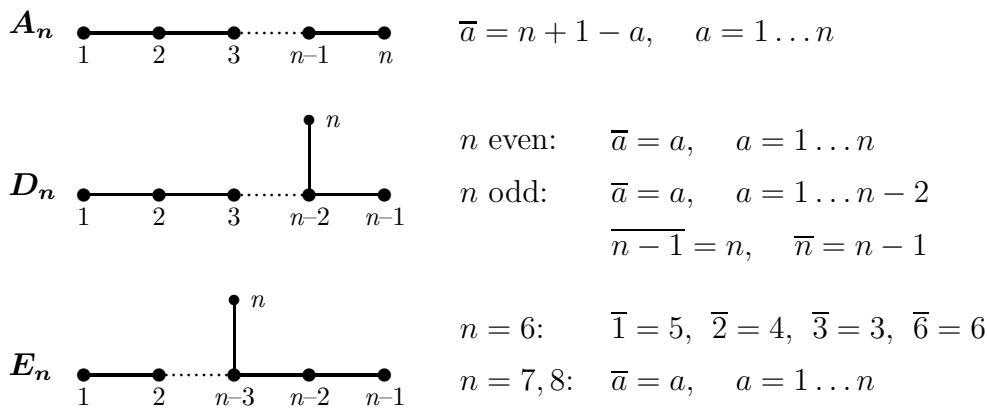
Dynkin diagrams of the simply-laced
Lie algebras. The numbers show our labelling convention for the nodes. The 
explicit form of the conjugation that appear in eq.~\rf{PeriodY} has 
also been included.}
\label{DynkinDiag}
}

\sect{The $\bfm{\mu\rightarrow-\mu}$ continuation of the TBA equations.}
\label{TBA}

~\indent
We are now in a position to give the TBA argument of 
\cite{DDT,DDTunpub}, relating
the changes from massive into massless magnonic TBA
systems to the continuation $\mu\rightarrow -\mu$ of the
corresponding actions. Consider the massive magnonic TBA system
specified by~\rf{EnTMassive}, and assume that it corresponds to some
two-dimensional action of the form~\rf{Action}.
On dimensional grounds, the coupling constant $\mu$ is related to 
the mass scale $M$ as
\begin{equation}
\mu= \kappa\> M^{\>\frac{2}{P}}\>,
\lab{LambdaM}
\end{equation}
with $\kappa$ a dimensionless (non-perturbative) constant.
Correspondingly, the dimensionless function $F_0(r)=RE_0(R)/2\pi$ is
expected to be a regular function of $r^{\frac{2}{P}}$, which
suggests that the TBA system of the same theory with $\mu\rightarrow
-\mu$ can be obtained by putting $r=MR$ on the ray
\begin{equation}
r=\ex{i\>\frac{\pi P}{2}}\> \rho\>, \qquad \rho\in{\fl R}^+\>,
\lab{Continuation}
\end{equation}
where $\rho$ is the dimensionless overall scale of the resulting
theory. Notice that this transformation makes the energy
terms~\rf{EnTMassive} complex. However, the explicit calculations
presented in~\cite{PATROB} for the scaling Lee-Yang model support the
expectation that the ground-state scaling function $F_0(r)$ evaluated
at $r=\ex{i\>\frac{\pi P}{2}}\> \rho$ is real up to some value
$\rho_0$ where it exhibits a branch point. And, moreover, that its
value indeed corresponds to the ground state of the theory with
$\mu\rightarrow -\mu$ for $\rho=|r|<\rho_0$. In the following, we
will assume that a similar result holds in general.

The idea of \cite{DDT,DDTunpub} is to consider the
massive magnonic TBA system specified by~\rf{EnTMassive}, and to
study the effect of~\rf{Continuation} on the
corresponding Y-system, in the spirit of \cite{PATROB,PATROB2}. 
The analytically
continued Y-functions will also be solutions to the original
Y-system, but with a different asymptotic behaviour. In the
following, it will be useful to display the dependence of the
solutions to the Y-system on $r$: $Y_a^i(\theta)\equiv
Y_a^i(r,\theta)$. Then, the form of the transformed energy terms
implies that the analytically continued Y-functions have the
following real-valued asymptotic behaviour
\begin{eqnarray}
&&Y_a^i\left(\ex{i\>\frac{\pi P}{2}} \rho, \theta+i\>\frac{\pi P}{2}\right)
\; \buildrel \theta\;\rightarrow -\infty \over{\hbox to
50pt{\rightarrowfill}}
\begin{cases}
\exp\Bigl(\frac{1}{2}\> \mu_a \rho \>\ex{-\theta}\Bigr)& \text{for}\quad 
i=l, \lab{RightAnalytic}\\
y_a^i& \text{for}\quad i\not=l,
\end{cases}\\[5pt]
&&Y_a^i\left(\ex{i\>\frac{\pi P}{2}} \rho, \theta-i\>\frac{\pi P}{2}\right)
\; \buildrel \theta\;\rightarrow +\infty \over{\hbox to
50pt{\rightarrowfill}}
\begin{cases}
\exp\Bigl(\frac{1}{2}\> \mu_a \rho \>\ex{+\theta}\Bigr)& \text{for}\quad 
i=l, \\
y_a^i& \text{for}\quad i\not=l,
\end{cases}
\end{eqnarray}
and, taking the periodicity~\rf{PeriodY} into account,
\begin{eqnarray}
&&Y_a^i\left(\ex{i\>\frac{\pi P}{2}} \rho, \theta+i\>\frac{\pi P}{2}\right)
= Y_{\>\overline{a}}^{\>\overline{\imath}}\left(\ex{i\>\frac{\pi P}{2}} 
\rho, \theta-i\>\frac{\pi P}{2}\right)\nonumber \\
&&
\hspace{4cm}\; \buildrel \theta\;\rightarrow +\infty \over{\hbox to
50pt{\rightarrowfill}}
\begin{cases}
\exp\Bigl(\frac{1}{2}\> \mu_a \rho \>\ex{+\theta}\Bigr)& \text{for}\quad 
i=\overline{l}, \\
y_a^i& \text{for}\quad i\not=\overline{l},
\end{cases}
\lab{LeftAnalytic}
\end{eqnarray}
where we have used that $\overline{\overline{l}}=l$, and that 
$\mu_{\overline{a}}=\mu_a$.

If we compare~\rf{RightAnalytic} and~\rf{LeftAnalytic} 
with~\rf{MasslessAsymp}, we observe  that
\begin{equation}
\widetilde{Y}_a^i(\rho,\theta)=Y_a^i\left(\ex{i\>\frac{\pi P}{2}}\rho, 
\theta+i\>\frac{\pi P}{2}\right)\equiv 
{\rm e}^{\>\widetilde{\varepsilon}_a^{\>i}(\rho,\theta)}
\end{equation}
provides a solution to the massless magnonic TBA system specified by the 
energy terms
\begin{equation}
\nu_a^i(\theta)= \delta_{i,l}\> \frac{1}{2}\> \mu_a\> \rho\> 
{\rm e}^{-\theta} + \delta_{i,\overline{l}}\> \frac{1}{2}\> \mu_a\> 
\rho\> {\rm e}^{+\theta}\>.
\lab{EnTMasslessB}
\end{equation}
In other words, the massive magnonic TBA system defined
by~\rf{EnTMassive} is related to the massless system specified
by~\rf{EnTMassless} by means of the continuation $\mu\rightarrow-\mu$
provided that the ${\fl Z}_2$ symmetry used in the construction of
the latter coincides with the conjugation that characterises the
periodicity conditions of the Y-functions; {\em i.e.\/},
$\omega(l)=\overline{l}$.{}~\footnote{
~The same arguments applied to the analytical continuation
$r\rightarrow \ex{i\>\pi P}\> r$ lead to
\begin{equation}
Y_a^i\left(\ex{i\>{\pi P}}\> r, 
\theta+i\>{\pi P}\right)= Y_a^i\left(r, \theta\right)\>,
\lab{Identity}
\end{equation}
which is an identity originally obtained in~\cite[eq.~2.8]{PATROB2}.
There, it was deduced as a consequence of the claim that
$Y_a^i\left(r, \theta\right)$ can be expanded as a power series in
the two variables $a_\pm=\left(r\ex{\pm \theta}\right)^\frac{1}{ P}$
with a finite domain of convergence about $a_+=a_-=0$.
Taking~\rf{LambdaM} into account, this analytic continuation corresponds 
to leaving $\mu$ invariant. }

\sect{$\bfm{\mu\rightarrow-\mu}$ continuation in the HSG models.}
\label{HSGTBA}

~\indent
An alternative understanding of the relation between the
continuation $\mu\rightarrow-\mu$ of the perturbed CFT action and the
transformation changing a massive magnonic TBA system into a massless
one can be obtained by considering the Lagrangian formulation of the
simply-laced Homogeneous sine-Gordon (HSG)
theories. This will be discussed in
the next section. But first, we clarify the relationship between
the HSG TBA equations and the magnonic TBA systems of~\cite{DKTBA}.

The Homogeneous sine-Gordon theories are integrable perturbations of
level~$k$ $G$-parafermions, that is of coset CFTs of the form
$G_k/U(1)^{r_G}$, where $G$ is a simple compact Lie group, $k>1$ is
an integer, and $r_G$ is the rank of $G$~\cite{QHSG}. In the
following, we will use $G$ to denote both the Lie group and the
Dynkin diagram of its Lie algebra, and we will restrict ourselves to
the case of simply-laced~$G$. The exact $S$-matrices of the simply-laced HSG theories have been constructed in~\cite{SMATHSG} (see
also~\cite{PATRICKTBA,SU3TBA}). They are always diagonal and describe
the scattering of a set of stable solitonic massive particles
labelled by two quantum numbers, $(i,a)$, where $i=1\ldots r_G$ and
$a=1\ldots k-1$. In other words, there is a stable particle for each
node of $G\times A_{k-1}$, the product of the Dynkin diagrams $G$ and
$A_{k-1}$. The mass of the particle $(i,a)$ is
\begin{equation}
M_a^i= M m_i \mu_a\>,
\end{equation}
where $M$ is a dimensionful overall mass scale, $m_1\ldots m_{r_G}$
are $r_G$ arbitrary non-vanishing relative masses, one for each node
of the Dynkin diagram $G$, and $\mu_a= \sin\left(\pi a/k\right)$ are
the components of the Perron-Frobenius eigenvector of the $A_{k-1}$
Cartan matrix. The $S$-matrix elements depend on a further set of
real resonance parameters $\sigma_{ij}=-\sigma_{ji}$ defined for each
pair $\{i,j\}$ of neighbouring nodes on $G$. They are most
conveniently specified by assigning a variable $\sigma_i$ to each
node of $G$ and setting $\sigma_{ij}=\sigma_i -\sigma_j$. The
resulting set of parameters $M$, $\{m_i\}$, and $\{\sigma_i\}$ is
redundant, but the obvious symmetries $M\rightarrow \alpha M$,
$\{m_i\rightarrow \alpha^{-1} m_i\}$, and $\{\sigma_i\rightarrow
\sigma_i+\beta\}$ ensure that there are $2r_G -1$ independent
adjustable parameters, which is one of the interesting features of
these theories.

The TBA equations of the HSG models have the standard form for a
diagonal scattering theory, although care is needed in their
derivation because parity is not a symmetry~\cite{SU3TBA}. There is a
pseudoenergy $\widehat{\epsilon}_a^{\>i}(\theta)$ for each of the
$(k-1)\times r_G$ stable particles, and the mass scales influence
them via $(k-1)\times r_G$ energy terms
\begin{equation}
\widehat{\nu}_a^{\>i}(\theta)= M_a^i R\cosh\theta = m_i \mu_a r\cosh\theta\>,
\lab{EnHSG}
\end{equation}
where $r=MR$. Using the same conventions of~\rf{TBAGen}, the 
pseudoenergies solve the TBA equations
\begin{equation}
\widehat{\nu}_a^{\>i}(\theta)= \widehat{\epsilon}_a^{\>i}(\theta)+
\sum_{b=1}^{k-1}\left(\phi_{ab}\ast \widehat{L}_b^{\>i} (\theta) -
\sum_{j=1}^{r_G}
\>G_{ij}\>
\psi_{ab}\ast \widehat{L}_b^{\>j} (\theta-\sigma_j+\sigma_i)
\right)\>.
\lab{TBAHSG}
\end{equation}
Then, the dimensionless effective central charge $c(r)$ is 
expressed in the usual way by~\rf{EffCC}, with $r_H=k-1$. 
Its limiting value as $r\rightarrow 0$ with all the other 
parameters fixed was calculated in~\cite{SU3TBA}, with the result
\begin{equation}
\lim_{r\rightarrow0} c(r)=\frac{k-1}{k+r_G}\> h_G\> r_G\>,
\end{equation}
which is the central charge of the $G_k/U(1)^{r_G}$ coset CFT. 
This holds for any fixed choice of the mass scales $0<m_i <+\infty$ 
and resonance parameters $-\infty<\sigma_i<+\infty$. 
Other exact multiple scaling limits, where the parameters $m_i$ and 
$\sigma_i$ approach particular limiting values while $r\rightarrow 0$, 
have been discussed in~\cite{PATRICKTBA}.
In the opposite, $r\rightarrow+\infty$, limit, $c(r)$ tends to 
zero, as expected for a massive theory.

In order to emphasise the similarities of the HSG TBA systems with the 
magnonic TBA systems of~\cite{DKTBA}, it is convenient to eliminate 
the explicit dependence of the TBA equations on the resonance parameters by 
writing them in terms of
\begin{equation}
\varepsilon_a^i(\theta)= \widehat{\varepsilon}_a^{\>i}(\theta-\sigma_i).
\end{equation}
Then,~\rf{TBAHSG} becomes
\begin{equation}
\nu_{a}^{\>i}(\theta)= \varepsilon_{a}^i (\theta)+
\sum_{b=1}^{k-1}\left(\phi_{ab}\ast L_{b}^i (\theta) -
\sum_{j=1}^{r_G}
\>G_{ij}\>
\psi_{ab}\ast L_{b}^j (\theta)
\right)\>,
\lab{TBAHSG2}
\end{equation}
where
\begin{equation}
\nu_a^i(\theta)=\widehat{\nu}_{a}^{\>i}(\theta-\sigma_i)= 
\frac{1}{2}\> m_i^+ \mu_a r \>\ex{-\theta} + 
\frac{1}{2}\> m_i^- \mu_a r \>\ex{+\theta}\>,
\lab{EnHSG2}
\end{equation}
and we have introduced
\begin{equation}
m_i^\pm = m_i\> \ex{\pm\sigma_i}\>.
\lab{Mplusminus}
\end{equation}
Eq.~\rf{TBAHSG2} is identical to~\rf{TBAGen} for $H=A_{k-1}$, which
exhibits that the HSG TBA systems corresponding to perturbations of
the $G/U(1)^{r_G}$ coset CFT are massive versions of the TBA systems
constructed in~\cite{DKTBA} in terms of the product $G\times
A_{k-1}$. The explicit asymmetric $\theta\rightarrow -\theta$
structure of the energy terms~\rf{EnHSG2} indicates that the HSG
theories are not parity symmetrical in general.

Formally, the magnonic $G\times A_{k-1}$ TBA systems of~\cite{DKTBA}
could be recovered from~\rf{TBAHSG2} by suitably choosing the
arbitrary parameters $m_i^+$ and $m_i^-$. Namely, $m_i^\pm=
\delta_{i,l}$ for the massive system specified by~\rf{EnTMassive},
and $m_i^-= \delta_{i,l}$ together with $m_i^+= \delta_{i,\omega(l)}$
for the massless system whose energy terms are~\rf{EnTMassless}.
However, since the HSG theories are purely massive these choices are
not permitted. In general, the connection between the massive HSG,
and the magnonic $G\times A_{k-1}$ TBA systems is recovered in a
different, more subtle, way: the latter are the effective TBA systems
describing the crossovers of the HSG theories for particular limiting
values of their parameters.
Crossover phenomena in the HSG theories were discussed in detail
in~\cite{PATRICKTBA}. One of the main results of that paper is that
the HSG theories exhibit a crossover at $r\approx 2/m_{pq}$ for each
`unshielded' mass scale $m_{pq}$ within the set of numbers given by
\begin{equation}
m_{pq}= \sqrt{m_p m_q}\>{\rm e}^{\> (\sigma_q-\sigma_p)/2} = 
\sqrt{m_q^+ m_p^-}\>, \qquad p,q=1\ldots r_G\>.
\lab{Crossover}
\end{equation}
Moreover, when the unshielded scale $m_{pq}$ is `well separated' 
from the others, the crossover at $r\approx 2/m_{pq}$ is described 
by an effective TBA system of the form~\rf{TBAHSG2} with energy terms
\begin{equation}
\nu_{a}^i(\theta)= 
\delta_{i,p}\>\frac{1}{2}\> m_{pq}\> \mu_a r \>\ex{-\theta} + 
\delta_{i,q}\>\frac{1}{2}\> m_{pq}\> \mu_a r \>\ex{+\theta}
\lab{EnHSG3}
\end{equation}
(see~\cite{PATRICKTBA} for details). For $p=q\equiv l$, 
and $p=\omega(q)\equiv l$ we recover~\rf{EnTMassive}, 
and~\rf{EnTMassless}, respectively.

The Homogeneous sine-Gordon theories describe integrable
perturbations of parafermionic theories defined by an action of the
form~\rf{Action}~\cite{QHSG,PATRICKTBA}, and we can investigate the
continuation $\mu\rightarrow-\mu$ using the methods of
section~\ref{TBA}. The HSG TBA equations~\rf{TBAHSG2} with the energy
terms~\rf{EnHSG2} provide solutions to the Y-system~\rf{YSystem}, for
$H=A_{k-1}$, with the asymptotic behaviour
\begin{eqnarray}
Y_a^i(r,\theta)
&&\hspace{-0.2cm}\buildrel \theta\;\rightarrow -\infty \over{\hbox to
50pt{\rightarrowfill}}
\exp\Bigl(\frac{1}{2}\> m_i^+\mu_a r \>\ex{-\theta}\Bigr)\nonumber\\[5pt]
&&\hspace{-0.2cm}\buildrel \theta\;\rightarrow +\infty \over{\hbox to
50pt{\rightarrowfill}}
\exp\Bigl(\frac{1}{2}\> m_i^-\mu_a r \>\ex{+\theta}\Bigr)\>.
\lab{HSGAsymp}
\end{eqnarray}
Consequently, they are completely characterised by the value of the parameters $m_i^+$ and $m_i^-$, for $i=1\ldots r_G$. 
Then, the results of section~\ref{TBA} imply that the continuation $\mu\rightarrow-\mu$ makes the Y-functions change according to
\begin{equation}
Y_a^i\left(r,\theta\right) \rightarrow \widetilde{Y}_a^i\left(r,\theta\right) = Y_a^i\left({\rm e\>}^{i\>\frac{\pi P}{2}}r,\theta+i\>\frac{\pi P}{2}\right)
= Y_{\overline{a}}^{\overline{\imath}}\left({\rm e\>}^{i\>\frac{\pi P}{2}}r,\theta-i\>\frac{\pi P}{2}\right)\>.
\end{equation}
In turn, taking~\rf{PeriodY} into account, the transformed Y-functions have the asymptotic behaviour
\begin{eqnarray}
\widetilde{Y}_a^i(r,\theta)
&&\hspace{-0.2cm}\buildrel \theta\;\rightarrow -\infty \over{\hbox to
50pt{\rightarrowfill}}
\exp\Bigl(\frac{1}{2}\> m_i^+\mu_a r \>\ex{-\theta}\Bigr)\nonumber\\[5pt]
&&\hspace{-0.2cm}\buildrel \theta\;\rightarrow +\infty \over{\hbox to
50pt{\rightarrowfill}}
\exp\Bigl(\frac{1}{2}\> m_{\overline\imath}^-\mu_a r \>\ex{+\theta}\Bigr)\>.
\lab{HSGAsymp2}
\end{eqnarray}
Therefore, we conclude that the $\mu\rightarrow -\mu$ continuation of the HSG TBA system with parameters $\{m_i^+,m_i^-\}$ is equivalent to the transformation
\begin{equation}
m_i^+\rightarrow \widetilde{m}_i^+=m_i^+ \>, \quad m_i^-\rightarrow \widetilde{m}_i^-=m_{\overline\imath}^-\>, \quad \forall\>i=1\ldots r_G\>.
\lab{HSGTBAEquiv}
\end{equation}

Taking into account the relationship between the massless and massive magnonic systems and the HSG TBA equations described in the paragraph around eq.~\rf{Crossover}, 
this generalises the connection between the continuation $\mu\rightarrow -\mu$ and the transformation between massless and massive systems originally pointed out in~\cite{DDT}, which was discussed in the previous section.
To be specific, let us assume that the 
scale $m_{pp}$ is `unshielded' and `well separated' from the others, so that the HSG theory exhibits a crossover at $r\approx 2/m_{pp}$. According to~\rf{EnHSG3}, the effective TBA system that describes this crossover is one of the {\em massive} magnonic TBA systems considered in section~\ref{Magnonic}. Then, since
\begin{equation}
m_{pp}=\sqrt{m_p^+ m_p^-} = \sqrt{m_p^+ \widetilde{m}_{\overline{p}}^-}\equiv\widetilde{m}_{p\overline{p}}
\end{equation}
the same theory with $\mu\rightarrow -\mu$ will exhibit a crossover at $r\approx 2/\widetilde{m}_{p\overline{p}}$, which is now effectively described by a {\em massless} magnonic TBA system.

\sect{The $\bfm{\mu\rightarrow-\mu}$ continuation and T-duality.}
\label{HSG}

~\indent
We will now show that the equivalence between the continuation $\mu\rightarrow -\mu$ and the transformation~\rf{HSGTBAEquiv} can be deduced in a completely different way using
the original Lagrangian formulation of the HSG theories in terms of a gauged WZW action modified by a potential. 

Let us denote by $g$ the Lie algebra of the group $G$. The theories corresponding to perturbations of the $G_k/U(1)^{r_G}$ coset CFT have 
actions~\cite{HSG,QHSG}
\begin{equation} 
S_{\rm HSG}[{\gamma},A_\pm]= k\Bigl(S_{\rm gWZW}[{\gamma},A_\pm]
\>-\int d^2x\> V({\gamma})\Bigr)\>.
\lab{ActionHSG}
\end{equation}
Here, ${\gamma}={\gamma}(t,x)$ is a
bosonic field that takes values in some faithful representation of the
compact Lie group $G$, and $A_\pm$ are non-dynamical gauge fields taking values in the Cartan subalgebra of $g$ associated with $H\simeq U(1)^{r_G}$, a maximal torus of $G$. Then, $kS_{\rm gWZW}$ is the
gauged  WZW action corresponding to the coset $G_k/H$.
The potential is
\begin{equation} 
V({\gamma}) =\frac{m_0^2}{4\pi} \> \langle
\Lambda_+ , {\gamma}^\dagger \Lambda_- {\gamma}\rangle\>,
\lab{Potential}
\end{equation}
where $m_0^2$ is a bare overall mass scale, $\langle\;,\>\rangle$ is the
Killing form of $g$, and
$\Lambda_\pm=i\bfm{\lambda}_\pm\cdot\bfm{h}$ are two arbitrary elements
in the same Cartan subalgebra of $g$ where $A_\pm$ take values. They are specified by two $r_G$-dimensional vectors $\bfm{\lambda}_+$ and
$\bfm{\lambda}_-$, and we will make this dependence explicit by writing
\begin{equation}
V \equiv V[\lav_+,\lav_-]\>.
\end{equation}

In~\cite{TDUAL}, it was shown that there is a group of T-duality
transformations that relate the HSG models corresponding to different
values of $\lav_\pm$. Namely, there is a duality transformation for
each Weyl transformation
$\sigma\in{\cal W}(G)$ that relates the models specified by the
following potentials
\begin{equation} V[\lav_+,\lav_-] 
\; \buildrel {\rm T-duality}\over{\hbox to 60pt{\rightarrowfill}} \;
V[\lav_+,\sigma(\lav_-)]\>,
\lab{TPot}
\end{equation}
and provides a map between two different phases of the model. To spell
this out, we have to be more precise about the possible values of
$\lav_+$ and $\lav_-$. The HSG theories are massive for any choice of
$\bfm{\lambda}_+$ and $\bfm{\lambda}_-$ such that
$\bfm{\lambda}_\pm\cdot \bfm{\alpha}\not=0$ for all the roots
$\bfm{\alpha}$ of $g$~\cite{HSG}. This makes possible to choose the
basis of simple roots of $g$, $\Delta=\{\bfm{\alpha}_1\ldots
\bfm{\alpha}_{r_G}\}$, such that $\bfm{\lambda}_+\cdot
\bfm{\alpha}_i>0$ for all $i=1\ldots r_G$. In other words, without
losing generality, we can restrict $\bfm{\lambda}_+$ to take values
inside $C(\Delta)$, the fundamental Weyl chamber with respect to
$\Delta$; {\em i.e.\/}, $\bfm{\lambda_+}\in C(\Lambda)$. Then, the
different phases of the theory are characterised by the domain where
$\bfm{\lambda}_-$ takes its values~\cite{TDUAL}. Since all the Weyl
chambers are permuted by ${\cal W}(G)$, there is a phase for each Weyl
transformation $\sigma\in {\cal W}(G)$ corresponding to
$\bfm{\lambda}_-\in \sigma^{-1}\left(C(\Delta)\right)$, which
justifies the interpretation of~\rf{TPot} as a map between two
different phases.

In the semiclassical limit, the vectors $\lav_+$ and $\lav_-$ are 
related to the TBA parameters $\{m_i,\sigma_i\}$. Consider an 
arbitrary phase where 
$\bfm{\lambda}_-\in \sigma^{-1}\left(C(\Delta)\right)$. Then, 
\begin{equation}
\sigma(\lav_-)=\lav_-^{\bullet}\in C(\Delta)\>.
\end{equation}
and the relationship is as follows~\cite{HSG,PATRICKTBA}
\begin{equation}
\lav_+ =\sum_{i=1}^{r_G} m_i^+ \> \lav_i\>, \qquad
\lav_-^\bullet =\sum_{i=1}^{r_G} m_i^- \> \lav_i\>,
\end{equation}
where $m_i^\pm$ are the semiclassical counterparts of the parameters
defined in~\rf{Mplusminus}, and $\lav_i$, $i=1\ldots r_G$, are the
fundamental weights of $g$ that satisfy
$\lav_i\cdot\alv_j=\delta_{ij}$. Notice that all the choices of
$\lav_-$ related by the T-duality transformations share the same
$\lav_-^\bullet$. This is so because they have the same masses and
resonance parameters, which can be seen as a semiclassical
confirmation of the duality symmetry.

We can now easily describe the meaning of the continuation
$\mu\rightarrow-\mu$ in this approach. In~\rf{ActionHSG}, $S_{\rm
HSG}$ is a Lagrangian action defined on 1+1 Minkowski space. This is
in contrast with~\rf{Action}, which defines the model as a perturbed
conformal field theory in two-dimensional Euclidean space with the
role of the potential $kV(\gamma)$ being taken by the perturbing
operator $\mu\Phi$. Therefore, in this framework, the continuation
$\mu\rightarrow-\mu$ of~\rf{Action} corresponds to
$V[\lav_+,\lav_-]\rightarrow -V[\lav_+,\lav_-]$. Since
\begin{equation}
-V[\lav_+,\lav_-]=V[\lav_+,-\lav_-]\>,
\lab{PotentialSign}
\end{equation}
this amounts to changing the phase of the model by means of 
$\lav_-\rightarrow -\lav_-$.

Let us consider a generic phase 
where $\sigma(\lav_-)=\lav_-^\bullet\in C(\Delta)$.
In order to find out the effect of~\rf{PotentialSign} on the TBA parameters, we have to look for a Weyl transformation that takes $\sigma(-\lav_-)$ back into $C(\Delta)$; {\em i.e.\/}, we have to look for $\sigma_0\in {\cal W}(G)$ such that 
\begin{equation}
\sigma_0\sigma(-\lav_-) =-\sigma_0(\lav_-^\bullet) \in C(\Delta) 
\end{equation}
for any $\lav_-^\bullet \in C(\Delta)$. There is a unique Weyl transformation with this property: the so-called `longest Weyl group element' . 
Its explicit form can be found, for instance, in~\cite[appendix E.13]{CORN}. It can be written in a concise way via
\begin{equation}
\sigma_0 = -\omega_0\>,
\end{equation}
where
\begin{equation}
\omega_0=\begin{cases}
1\>, \quad\text{for}\quad g=a_1,\>d_{2n},\>e_7,\>e_8,\>b_n,\>c_n,\>f_4,\>g_2,& \\[5pt]
\text{Dynkin diagram automorphism}\>, \quad\text{otherwise}.
\end{cases}
\end{equation}
Notice that $\omega_0$ acts on the Dynkin diagram $G$. In particular for simply-laced~$G$,
\begin{equation}
\omega_0(i) = \overline{\imath} \quad \forall\> i=1\ldots r_G\>,
\end{equation}
where $\overline{\imath}$ is the conjugation defined in fig.~\ref{DynkinDiag}.

Therefore, we conclude that the continuation $\mu\rightarrow -\mu$ of the HSG action is equivalent to the transformation $\lav_-^\bullet \rightarrow \omega_0(\lav_-^\bullet)$. This amounts to
\begin{equation}
m_i^+\rightarrow m_i^+ \>, \quad m_i^-\rightarrow m_{\overline\imath}^-\>, \quad \forall\>i=1\ldots r_G\>,
\lab{HSGDUALEquiv}
\end{equation}
which coincides with the equivalence~\rf{HSGTBAEquiv},
deduced above using the Y-system continuation argument of
\cite{DDT,DDTunpub}.
The agreement between the two methods used to deduce~\rf{HSGTBAEquiv} and~\rf{HSGDUALEquiv} is one of the main results of this letter. 

\sect{Conclusions}
\label{Conclusions}

{}~\indent
The main result of this letter is the derivation of a new Lagrangian
interpretation of the relationship between the transformation of
massive into massless magnonic TBA systems, and the change of sign of
the coupling constant in the corresponding perturbed conformal field
theory actions.
This relationship was originally noticed by Al.~Zamolodchikov in the
study of the perturbation of the unitary minimal models by their
least relevant primary field~\cite{ZAMtric}, and it was found to be
true in many other cases. Its extent was clarified in~\cite{DDT,DDTunpub},
where it was related to the properties of the associated Y-system
by means of the study of the analytic continuation of the TBA
equations.

The novel interpretation arises in the context of the Homogeneous
sine-Gordon (HSG) theories~\cite{HSG,QHSG,SMATHSG}, whose TBA
equations are purely massive generalisations of the magnonic TBA
systems corresponding to products of Dynkin diagrams of the form
$G\times A_{k}$.
They  depend on $2\> {\rm rank}(G)-1$ independent adjustable
parameters, and the usual magnonic massive and massless systems are
recovered as the effective TBA systems that describe the crossovers
of these theories for particular limiting values of those
parameters~\cite{PATRICKTBA}. The HSG theories admit a Lagrangian
formulation in terms of a gauged Wess-Zumino-Witten action with a
potential term that takes the role of the perturbing operator in
their interpretation as perturbed conformal field theories.
In this framework, the relationship arises as a consequence of the
target-space (T-space) duality symmetries of the Lagrangian
action~\cite{TDUAL}. To be specific, changing the sign of the
coupling constant corresponds to an overall change of the sign of the
potential, which is T-dual to a transformation among the adjustable
parameters. This transformation changes the pattern of crossovers
exhibited by the theory in such a way that all the crossovers
effectively described by
massive magnonic TBA systems turn out to be described by massless
ones and vice versa, thus giving support to the claim that the
observed relationship can indeed be understood as a consequence of
T-duality.

Moreover, we have shown that the transformation among the adjustable
parameters, summarised by~\rf{HSGTBAEquiv}  and~\rf{HSGDUALEquiv},
can also be obtained through the generalisation of the arguments
of~\cite{DDT,DDTunpub} to study the analytic continuation of the HSG
TBA system. This is rather remarkable because the two methods used to
deduce it are expected to be valid in different regimes. Namely,
eq.~\rf{HSGTBAEquiv} was derived from the analytical continuation of
the HSG TBA system, making use of the properties of the corresponding
Y-systems. This is expected to hold for values of the dimensionless
overall scale $r$ smaller than some upper value $r_0$, which means
that we should be close enough to the UV limit.
In contrast, the Lagrangian arguments leading to~\rf{HSGDUALEquiv}
should be valid in the semiclassical, large~$k$, limit. The agreement
of the resulting transformations provides a second interpretation of
our results, as a non-perturbative check, enabled by integrability, of the semiclassical
Lagrangian arguments used in~\cite{TDUAL,GOMES} to study T-duality in a particular family of massive theories (for a different class of models, integrability was previously used for similar purposes in~\cite{FORGACS}).
Conversely, we expect that T-duality symmetries will provide relevant
information
to understand better the nature of the flows in the HSG models, and
the relationship between their TBA and Lagrangian parameters.


\acknowledgments{
We would like to thank Clare Dunning,
Francesco Ravanini and Roberto Tateo for
helpful discussions, and collaborations on many closely-related
topics.
JLM thanks Bologna (INFN and University) for hospitality.
This work was partly supported by the EC
network ``EUCLID", contract number HPRN-CT-2002-00325, and partly by
a NATO grant PST.CLG.980424.
JLM also thanks MEC (Spain) and FEDER (FPA2005-00188 and 
FPA2005-01963), and Incentivos from Xunta de Galicia
for financial support.}


\end{document}